\documentclass
[twocolumn,showpacs,preprintnumbers,amsmath,amssymb,prc]{revtex4}
\usepackage{graphicx}
\usepackage{dcolumn}
\usepackage{bm}
\usepackage{color}
\begin{document}
\title{Event simulations in a transport model for intermediate energy heavy ion
collisions: Applications to multiplicity distributions}
\author{S. Mallik$^1$, S. Das Gupta$^2$ and G. Chaudhuri$^{1}$}

\affiliation{$^1$Theoretical Physics Division, Variable Energy Cyclotron Centre, 1/AF Bidhan Nagar, Kolkata 700064, India}
\affiliation{$^2$Physics Department, McGill University, Montr{\'e}al, Canada H3A 2T8}


\begin{abstract}
We perform transport model calculations for central collisions of mass 120 on
mass 120 at laboratory beam energy in the range 20 MeV/nucleon to
200 MeV/nucleon.  A simplified yet accurate method allows calculation of
fluctuations in systems much larger than what was considered feasible in
a well-known and already existing model.  The calculations produce clusters.
The distribution of clusters is remarkably similar to that obtained in
equilibrium statistical model.
\end{abstract}


\maketitle

\section{Introduction}
There is an enormous amount of experimental and theoretical work on
multifragmentation in heavy ion collisions at intermediate energy.
There are two classes of theoretical models: (a) dynamical and
(b) statistical
where one assumes that because of two-body collisions the colliding systems
equilibrate and break up into many fragments according to availability of
phase space.  Unquestionably the most used and popular dynamical model
is the Boltzmann-Uehling-Uhlenbeck (BUU) model (also called by various other
names).  Here we restrict ourselves to BUU \cite{Bertsch}.  In the
original formulation, the BUU model gave an account of one body properties
\cite{Bertsch2} and thus was not suitable to describe multifragmantation.
Later it was extended to include fluctuations which made it suitable for
event by event simulation \cite{Bauer}.  This is the focus of our attention.
In the past event by event simulation in this model was limited to about mass
number 30 on 30.
The problem was a practical one, namely, it required a large computing effort.
We show that with a slight reformulation without changing any physics or
numerical accuracy we can very
significantly reduce the execution time and we can handle much larger
systems.  Computation becomes as short as an ordinary BUU calculation.
It is instructive to do large systems (finite number effects often hide
important bulk effects) and more importantly, the fragmentation must be
investigated over an energy range to unravel many interesting effects.
The objective of doing examples is to demonstrate that many revealing features
are seen. These also allow us to relate with other models.\\
There are many models
for multifragmentation.  There are some which can be labeled as
``quantum molecular dynamics'' type \cite{Aichelin1,Beauvais}
These are different in spirit to the model used here.  Closer in spirit
yet quite distinct are some studies based on a Langevin model
\cite{Ayik,Randrup,Chomaz,Rizzo,Napolitani} where
we have mentioned only a few.  We will have occasion to refer very briefly
to only a small number of these.  The literature in the Langevin approach
is huge.

\section{The Prescription}

The basic features of our transport model calculation are contained in
the Boltzmann-Uehling-Uhlenbeck (BUU)
model as developed in \cite{Bertsch} and \cite{Bauer} but some modifications
were made.  For brevity we will almost entirely skip the physical motivations
and details
for the models of refs. \cite{Bertsch} and \cite{Bauer} as they are not only adequately discussed
in the original papers but also elsewhere \cite{Rizzo,Chomaz} where some
different models are also introduced.
The modifications we make to refs. \cite{Bertsch} and \cite{Bauer} are
discussed fully.

The start of our consideration is the cascade model \cite{Cugnon}.  Here each
nucleus is considered
as a collection of point nucleons whose positions are assigned by Monte-Carlo
sampling.  The projectile nucleus $A$ approaches the target $B$ with a beam
velocity and two body collisions between the nucleons take place.  When these
finish we have one event.  We only consider $B$ same as $A$ and central
collisions.  It is convenient to run several events
simultaneously.  Let us label the number of runs by $\tilde{N}$.  In cascade
the different runs do not communicate with each other.  Thus nucleus 1 hits
nucleus 1', nucleus 2 hits nucleus 2',....nucleus $\tilde{N}$ hits nucleus
$\tilde{N}$'.  In BUU we introduce communication between runs.  What we were
calling nucleons we now call test particles (abbreviated from now on as $tp$).
The density $\rho(\vec{r})$ is given by $n/(\delta r)^3\tilde{N}$
where $n$ is the number of $tp$s in a small volume $(\delta r)^3$.  As far as
collisions go, in usual applications of BUU one still segregates different
runs.  By segregating the collisions
one is able to use $\sigma_{nn}$, the nucleon-nucleon cross-section and
reduce computation.  If we considered collisions between all $tp$'s, the
collision cross-sections would have to be reduced.  In between collisions
$tp$s move in a mean field (Vlasov propagation).
Applications of BUU as
summarised above have met much success in explaining average properties
such as average collective flows etc.

To explain multifragmentation, multiplicity $n_a$ as a function of $a$
where $a$ is the mass number of the composite,
one needs an event by event computation in the transport model.
Bauer et al made the following prescription \cite{Bauer}.  Now all $tp$s are
allowed to
collide with one another with a cross-section of $\sigma_{nn}/\tilde{N}$.
Collisions are further suppressed by a factor $\tilde{N}$ but when two
$tp$s collide not only those two but 2$(\tilde{N}-1)$ $tp$s contiguous
in phase space change momenta also.  Physically it represents two actual
particles colliding.  When collisions cease we have one event.  A second event
needs a new Monte-Carlo of $tp$s and then the evolution in time.

The prescription we use here is the following.  This is the middle ground
between ref. \cite{Bertsch} and ref. \cite{Bauer}.  As in ref. \cite{Bertsch} for nucleon-nucleon collisions we
consider 1 on 1'(event1), 2 on 2'(event2) etc. with cross-section $\sigma_{nn}$.
For event 1 we will consider $nn$ collisions only between 1 and 1'.
The collision is checked for Pauli blocking as in ref. \cite{Bertsch}.
If a collision
between $i$ and $j$ in event 1 is allowed we follow ref. \cite{Bauer} and pick
$\tilde{N}$
-1 $tp$s from all the $tp$s closest to $i$ and give them the same momentum
change $\Delta \vec{p}$ as ascribed to $i$.  Similarly we pick
$\tilde{N}$-1  $tp$'s closest to $j$ and ascribe them the momentum change
$-\Delta \vec{p}$, the same as suffered by $j$.  We will return to more
details about this later.  As a function of time this is continued till event 1
is over.  For event 2 we return to time $t$=0,
the original situation (or a new Monte-carlo
sampling for the  original nuclei), follow the above procedure but consider
$nn$ collisions only between 2 and 2'.  This can be repeated for as many events
as one needs to build up enough statistics.The advantage of this over
that used in ref. \cite{Bauer} is that here, for one event, $nn$ collisions need to be
considered between $(N_A+N_B)$ nucleons ($N_A$=number of nucleons in $A$,
$N_B$=number of nucleons in $B$) whereas in method of ref. \cite{Bauer},
collisions need
to be checked between $(N_A+N_B)\times\tilde{N}$ $tp$s. Hence, in our calculation, total number of combinations for two-body collision is reduced by a factor of 1/$\tilde{N}^2$. Since typically
$\tilde{N}$ is of the order of 100 this is a huge
saving in computation and has allowed
us to treat mass as large as 120 on 120 over a substantial energy range.
It is expected that the model used in ref. \cite{Bauer} and the one used
here will give similar results.
The number of collisions for one event should be about the
same in both prescriptions.  The characteristics of scattering are the same.
The objects that collide in our calculation arise from a coarse grain
representation
of the initial phase space population of two nuclei.  In ref. \cite{Bauer} a fine grain
representation is used.  But since many events are generated any difference should disappear.
The Vlasov propagation is the same.
For mass 40 on 40 we compare our results with those using the method of
ref. \cite{Bauer}(Fig.1).  The agreement between
the two calculations for multiplicities is remarkable.  We regard our method
as a very convenient short cut to the numerical modeling of ref. \cite{Bauer}.  The
theoretical formulation in ref. \cite{Bauer} is more appealing and ``democratic''
but numerically our method gives indistinguishable results.

One bonus of our prescription is that one sees some common ground between
the BUU approach and the ``quantum molecular dynamics'' approach.  In the
latter nucleons are represented by Gaussians in phase space; the centroids
have an $\vec{r}$ and a $\vec{p}$ which are originally generated by Monte-Carlo calculations.
These collide.  This corresponds to ``nucleons'' colliding in our prescription.
As the centroids move after collision, they drag the Gaussians
along.  The Gaussian wave packets in position and momentum space
provide the mean-field and Pauli blocking.  The Gaussians do not change their
shapes or widths.  These are very strong restriction and lead to very
different mean field propagation.
The Vlasov propagation
has much more flexibility  and originates from more fundamental theory.
\begin{figure}
\begin{center}
\includegraphics[width=\columnwidth,keepaspectratio=true]{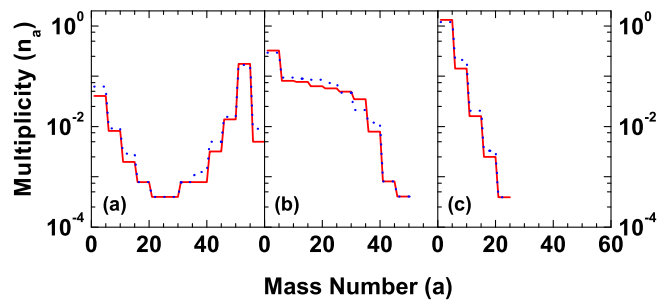}
\caption{(Color online) Comparison of mass distribution  calculated according to the prescription of ref. \cite{Bauer} (blue dotted lines) and the present work (red solid lines). The average value of 5 mass units is shown. The cases are for central collision of mass 40 on mass 40 for different beam energies (a) 25, (b) 50 and (c) 100 MeV/nucleon. 500 events were chosen at each energy.}
\end{center}
\end{figure}

\section{Some details of the simulations}
We provide some details of the calculation.  For Vlasov propagation we use the
lattice Hamiltonian method \cite{Lenk} of Lenk and Pandharipande which
accurately
conserves energy and momentum.  The mean field Hamiltonian for Vlasov
propagation is also adopted from that work.  The potential energy density is:
\begin{equation}
v(\rho(\vec{r}))=\frac{A}{2}\rho^2(\vec{r})+\frac{B}{\sigma +1}\rho^{\sigma+1}
(\vec{r})+\frac{c\rho_0^{1/3}}{2}\frac{\rho(\vec{r})}{\rho_0}\nabla_r^2
[\frac{\rho(\vec{r})}{\rho_0}]
\end{equation}
The values of the constants are $A$=-2230.0 MeV$fm^3$, $B$=2577.85 MeV$fm^{7/2}$,
$\sigma$=7/6, $\rho_0$=0.16$fm^{-3}$, $c$=-6.5 MeV$fm^{5/2}$.  The last term in the right hand side
of eq.(1) gives rise to
surface energy in finite nuclei. That favours the formation of larger
composites,
for example, the occurrence of a nucleus of $A$
nucleons over the formation of two nuclei of $A/2$ nucleons.  Entropy
works the other way.

\begin{figure}[b]
\begin{center}
\includegraphics[width=\columnwidth,keepaspectratio=true]{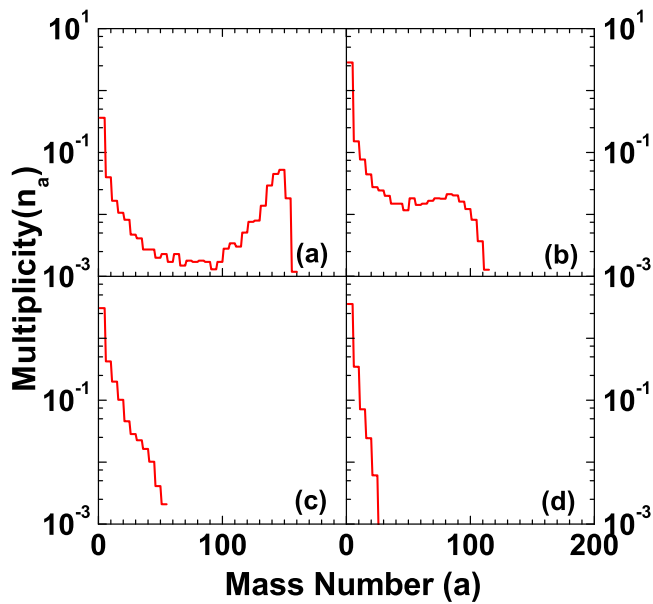}
\caption{(Color online) Mass distribution from BUU model calculation for $N_A=120$ on $N_B=120$ reaction
at beam energies (a)50 MeV/nucleon, (b)75 MeV/nucleon (c)100 MeV/nucleon and
(d)150 MeV/nucleon.  The average value of 5 mass units are shown. At each
energy 1000 events are chosen. Only central collisions are considered here but even at $E_p=50$ MeV/nucleon, nucleons in the peripheral region passes through and largest fragment remaining is less than the sum of the masses of the two nuclei.}
\end{center}
\end{figure}

\begin{figure}[t]
\begin{center}
\includegraphics[width=\columnwidth,keepaspectratio=true]{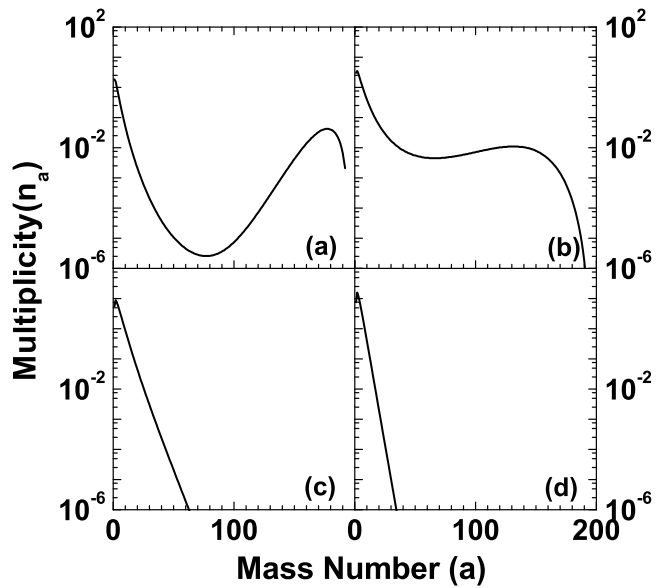}
\caption{Mass distribution from the Canonical Thermodynamical
Model (CTM) calculation for fragmentation of a system of mass $A_0=192$ at temperature (a)6.5 MeV, (b)7.5 MeV (c)10 MeV and
(d)14 MeV.}
\end{center}
\end{figure}

\begin{figure*}
\begin{center}
\includegraphics[width=11cm,keepaspectratio=true]{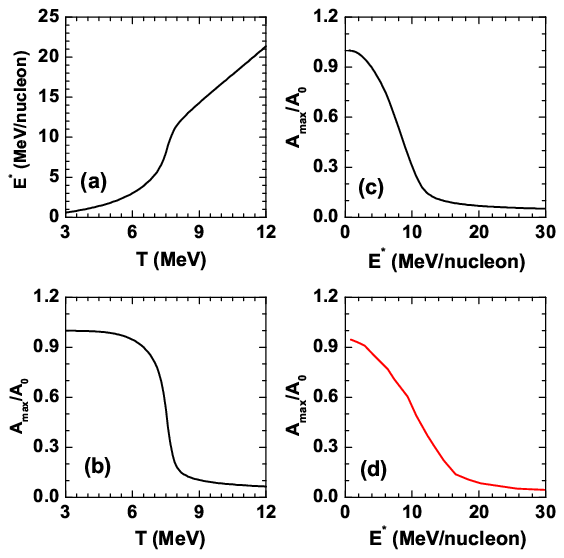}
\caption{ (Color online) Top left curve (a) is a Canonical Thermodynamical
Model (CTM) calculation for excitation ($E^*$) vs. temperature($T$) for $A_0=192$. Between 6 MeV and 7.5
MeV temperatures, $E^*$ rises quickly. The $dE^*/dT$ slope increases sharply
with mass size $A_0$ and is indicative of first order phase transition.
Bottom left curve (b) is also a CTM curve showing that the size of largest
cluster drops sharply between 6 MeV and 7.5 MeV. Again this is first order
liquid gas phase transition. Top right (c) is also with CTM but $A_{max}/A_0$
is plotted against excitation energy per nucleon instead of temperature. The
change of liquid to gas is necessarily slower, the range of energy for the
change is dictated by latent heat. Bottom right (d) is the calculation from
transport model.}
\end{center}
\end{figure*}

Further detail;s are :

(1) Calculations were done in a 200$\times 200\times 200 fm^3$ box. The configuration space was divided into  $1fm^3$ boxes.

(2) For results shown here the code was run from $t=0 fm/c$ to $t=200 fm/c$.  Positions and
momenta of $tp$s were updated every $\Delta t=0.3 fm/c$.

(3) For nucleon-nucleon collision we follow Appendix. B. of ref. \cite{Bertsch}.

(4) The number $\tilde{N}$ was set at 100.

(5) Once the two-body collisions are nearly over, contiguous boxes with $tp$s
that propagate together for a long time are considered to be part of the same
cluster. The contiguous boxes have at least one common surface and the nuclear density exceeds a minimum value ($d_{min}$). Different $d_{min}$ values as 0.002, 0.005, 0.01, 0.015 and 0.02 fm$^{-3}$ are tried to check the sensitivity of this parameter. It is observed that the fragment multiplicity distribution is not changing very much with $d_{min}$, therefore we use $d_{min}=$0.01 fm$^{-3}$ for further calculations.

\section{Results}
In Fig.2 we show plots of multiplicity against mass number $a$ for 120 on 120.
Four beam energies are shown.  For each energy 1000 events were taken.
We show results of averages for groups of five consecutive mass numbers.
The  prominent feature we wish to point out is that at low beam energy
(50 MeV/nucleon) the multiplicity first falls with mass number $a$, reaches
a minimum, then rises, reaches a maximum before disappearing.  As the beam
energy increases the height of the second maximum decreases.  At beam
energy 75 MeV/nucleon the second maximum is still there but barely.  At higher energy
the multiplicity is monotonically decreasing, the slope becoming steeper as
the beam energy increases. This evolution of shape of the multiplicity
distribution is of significance as we will emphasize soon but let us
point out this evolution of shape was a long time
prediction of the canonical thermodynamic model (CTM) \cite{Dasgupta,Das}.
For transport models the natural variable is the beam energy.  For CTM
the natural variable is the temperature $T$.  For illustration we have shown
the multiplicity distribution for a system of 192 particles in CTM at
temperatures of 6.5 MeV, 7.5 MeV, 10 MeV and 14 MeV (Fig. 3).  The calculations with BUU
and CTM are so different that the similarity in the evolution of the shape
in multiplicity distribution is very striking.  Indeed this correspondence
provides the support for assumptions of statistical model from a microscopic
calculation.

To proceed further with the correspondence between the two models we need to
establish a connection between $E_p$ of BUU and temperature $T$
of CTM.  Temperature $T$ of CTM will give an average excitation energy
$E^*$ of the multifragmenting system in its center of mass ($cm$) \cite{Das}.
We can calculate the excitation energy ($E^*$) in the $cm$ from
($E_p$) by direct kinematics by assuming that the projectile and the
target fuse together.  In that case the excitation energy is
$E^*=A_pE_p/(A_p+A_t)$ where $A_p$ and $A_t$ are projectile and target
masses respectively.  This value is too high as a measure of the excitation
energy of the system which multifragments.  The nucleons at the edges
of the two nuclei pass through carrying a lot of energy and are not part
of the multifragmenting system.  These are pre-equilibrium particles.
In experiments
about 20$\%$ of nucleons are emitted as fast pre-equilibrium particles:
see for example \cite{Xu,Frankland}.
Further details can be found in ref. \cite{Mallik11} but this is what we
do basically.  We go to the cm of the two ions to do BUU and at the end
discard 20$\%$
particles (these have the highest energies), and measure energy (potential
plus kinetic) of the rest.  To find the excitation energy we subtract
Thomas-Fermi ground state energy of the rest with the Hamiltonian of eq.(1).

Fig.4 gives some CTM results and also makes a comparison of one CTM result
with transport model result.  The top left
diagram is $E^*$ vs.$T$ in CTM for 192 particles ($A_0$=192=80$\%$ of 240).
This approximates usual $E^*$ vs $T$ for first order phase transition.
There is a boiling point temperature $T$ which remains constant as energy
increases.  In the example here
because we have a very finite
system, the slope $dE^*/dT$ is not infinite but high.  Let us now consider
lower left diagram again drawn in CTM.  Here $A_{max}$ is the average value of
the largest
cluster.  A high value of $A_{max}/A_0$ means liquid phase and low values
means gas phase.  The criteria of deciding which composites belong to the
gas phase and which to the liquid phase
are discussed in detail in two previous papers \cite{Chaudhuri1,Chaudhuri2}.

In the bottom left diagram, one sees more dramatically that in a short
temperature interval
liquid has transformed into gas.  The only input in our transport model
is the beam energy.  The common dynamical variable in both our model
and CTM is $E^*$.  Of course the $E^*$ in CTM is an average whereas
the $E^*$ in transport model is a microcanonical $E^*$.
In the top right corner of Fig.4 is the plot of
$A_{max}/A_0$ as a function of $E^*$ in CTM.  The transformation from liquid
to gas is more gradual, essentially spanning the energy range across which,
liquid transforms totally into gas.  Even for a large system, where the
transformation of liquid to gas as a function of temperature is very abrupt,
the transformation as a function of energy per particle will be quite smooth.
The bottom right in Fig.4 is from our transport model calculation.
The similarity with the CTM graph is close enough that we conclude
the transport model calculation gives evidence of
liquid-gas phase transition. To find closer correspondence between transport model calculations and CTM, it will be best if we can deduce at least an approximate value of temperature for each beam energy.  For an interacting system this is very non-trivial. Formulae like $\frac{E^*}{A}=\frac{3T}{2}$ are obviously inappropriate. One might try to exploit the thermodynamic identity $T=(\frac{\partial E}{\partial S})_V$. This requires obtaining a value of the entropy for an interacting system. We will be working on this in future.

In concluding this section, we mention that while we have established a
correspondence between transport model results and CTM results, a more
natural choice would have been to compare transport model results with
multiplicity distributions obtained from the microcanonical statistical
multifragmentation model(SMM) \cite{Bondorf}.  These are not available to us.
However for the only cases investigated we found that CTM and SMM results were
quite close \cite{Botvina} so the correspondence we have found here between
transport model results and CTM will presumably hold for SMM also. Multiplicity distributions in $^{16}$O+$^{80}$Br were done with dynamic models before. The work in Ref. \cite{Souza} used molecular dynamics. The work in Ref. \cite{Leray} comes closer in spirit to ours. The colliding system was small and no attempt was made to link the work with statistical models or phase transition.

\section{Discussion}
We now look at one feature of the model that raised concerns and led to
a lot of work to propose alternative methods for calculations
\cite{Rizzo,Napolitani}.
This is related to dangers of crossing fermionic occupation limits
in the model here (as in the model of ref. \cite{Bauer}.).  As mentioned
already,
if Pauli blocking allows two $tp$s $i$ and $j$ to collide, then not only
these two but also $\tilde{N}-1 tp$s closest to $i$ and $\tilde{N}$-1
closest to $j$ move to represent that two actual nucleons scatter.  The $tp$s
that move with $i$ are denoted by $i_s$, with $s$=0 to $\tilde{N}-1$.The square
of the
distance is taken to be $d_{0s}^2=\frac{(\vec{r_{io}}-\vec{r_{is}})^2}{R^2}
+\frac{(\vec{p_{io}}-\vec{p_{is}})^2}{p_F^2}$.  Here $R$ is the radius of the
static nucleus of $A$=120
and $p_F$ the Fermi momentum..  The $tp$s $j_s$ are then chosen
from the rest of the $tp$s.  Define now $<\vec{p_i}>=\frac{\sum \vec{p_{is}}}
{\tilde{N}}$, similarly $<\vec{p_j}>$.  One then considers a collision
between $<\vec{p_i}>$ and $<\vec{p_j}>$ and obtain a $\Delta \vec{p}$ for
$<\vec{p_i}>$ and -$\Delta \vec{p}$ for $<\vec {p_j}>$.  This $\Delta \vec{p}$
is added to all $\vec {p_{is}}$ and $-\Delta {\vec{p}}$ to all $\vec{p_j}$.
Since the $tp$s are moved without verifying Pauli blocking there may
be cases where one exceeds the occupation limits for fermions.

\begin{figure}[h!]
\begin{center}
\includegraphics[width=8cm,keepaspectratio=true]{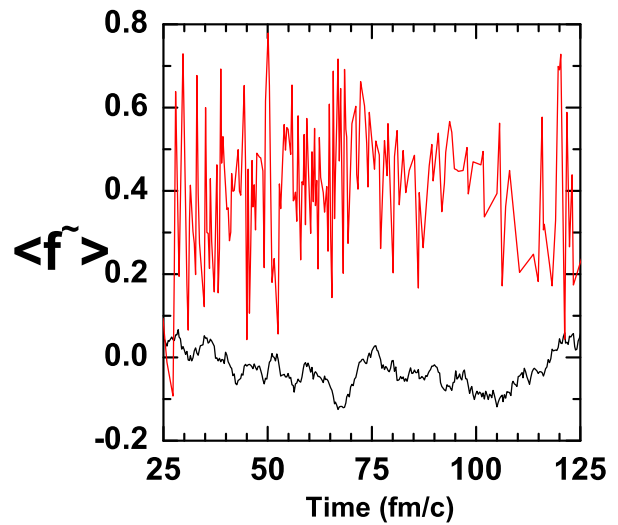}
\caption{ (Color online) Variation of average availability factor (see text)
with time (red line) for $N_A=120$ on $N_B=120$ reaction at beam energy 100
MeV/nucleon. The lower curve (black dotted line)is the average availability
factor $\langle \tilde{f} \rangle$ at the phase space points of arbitrarily chosen
120 test particles in an isolated mass 120 nucleus as they move in time. The
fluctuations from the value 0 reflects uncertainties, probably due to
fluctuation in initial Monte-Carlo simulations.}
\end{center}
\end{figure}

Initially the two ions have a very compact occupation at two different corners
of phase space.  Collisions make a far wider region of phase space available
to nucleons so this problem may not be severe.  An accurate estimation
of exceeding the fermionic limit of occupation at various parts of phase space
is very hard to compute in our
present problem but some measures are relevant.

For 120 on 120 at 100 MeV/nucleon beam energy (50 MeV/nucleon beam energy was studied also)
we follow one event as a function of time.  In every collision
in the event, $2\tilde{N}$
$tp$s change momenta.  To be specific, let the $tp$ $i_s$ move from
$\vec {r},\vec{p_{in}}$ to $\vec{r},\vec{p_{fi}}$.  We check if by moving $i_s$
to the final phase space point ($\vec{r},\vec{p_{fi}}$) we cross the fermionic
limit.  We build a six dimensional
unit cell in phase space around this final phase space
point \cite{Bertsch,Aichelin}.  The volume of the unit cell Should be
small so that one is investigating the phase-space occupation very near
$(\vec {r},\vec {p_{fi}})$, but it can not be too small since Monte-Carlo
simulation has noise which can cloud the actual effect.  In accordance with
past calculations we chose a volume of unit cell in phase space where
8 $tp$s is the maximum number allowed for fermions.
If $n$ is the number of $tp$s
(not including $i_s$) already in the unit cell we define an availability
factor $\tilde{f}=1.0-n/8$.  If $\tilde{f}$=0 we are already at the limit of
fermionic
occupation. If $\tilde{f}$ is negative we have crossed the quantum limit and
are in the
classical regime.  Any positive number between 0 and 1 will accommodate
additional
fermion.  For each collision there are 200 $\tilde{f}$'s to be calculated so
for each collision we get an ave $\tilde{f}$ and that is plotted in Fig.5.  We
have shown results  for $t$=25 to $125 fm/c$ when most of the action
takes place.  For reference we also plot average $\tilde{f}$ for randomly
chosen 120 $tp$s in a static mass 120 nucleus
as they move around in
time.  This number should ideally be 0 and not fluctuate.  The deviations
from zero in the static case probably largely arise
 due to fluctuations in Monte-Carlo sampling.
This degree of uncertainty must be also present in the values of $\tilde{f}$
we have plotted
for collisions.  In spite of these uncertainties the predominantly
positive values of $\tilde{f}$ as displayed  in Fig.5 lead us to believe that
the general trends we find in our calculation will hold.

If in a collision all of the 200 $tp$ moved to locations where $\tilde{f}$ were
all positive we will stay within the fermionic limits.  In case there is a $tp$
which does not satisfy this we can try to improve the situation by discarding
that $tp$ and choosing the next
available $tp$ to be part of the cloud.  Complications arise because
when some of the previously chosen $tp$s are discarded for new ones the
average momentum of the clouds will change, new $\Delta\vec{p}$ will
have to be used so the final resting spots obeying energy and momentum
conservation will change too.  An iterative procedure needs to be formulated but
convergence may be slow.

Alternative methods have been proposed.  The two papers which give procedural
details of moving two clouds of $tp$s from initial positions to final positions
with a stricter adherence to fermionic limits are refs.
\cite{Rizzo,Napolitani}.  Multiplicity distributions are not given so we
can not compare.  Even if the multiplicity distributions turn out to be similar,
higher order correlations can be very different.  The present work extended
the first proposed model of fluctuations in BUU to a larger system at many
energies and a very interesting lesson was learned.  The gross features of
multiplicity distribution do resemble strongly the results from equilibrium
statistical models which have proven very successful in explaining experimental data.
\section{Summary}
An event by event simulation of a transport model was made at collisions of moderately heavy ions at zero impact parameter. Multiplicity distributions were calculated. They are remarkably similar to those obtained from a equilibrium statistical model (CTM). This work therefore justifies the use of the equilibrium statistical model for data fitting. This statistical model implies first order phase transition in large nuclear systems at finite temperature. It will be of interest to quantify more precisely the correspondence of transport model result and statistical model results. That work is in progress.

\section{Acknowledgements}

Part of this calculation was performed in Variable Energy Cyclotron Centre
(VECC) and part at McGill University.  S. Das Gupta thanks the director of
VECC Dr. D. K. Srivastava for his strong support for the project and hospitality.
He also thanks Dr. A. K. Chaudhuri for hospitality.  S. Mallik thanks the
physics department of McGill University for a very productive and pleasant stay in Montreal. This work was supported in part by Natural Sciences and Engineering Research Council (NSERC) of Canada.


\begin{thebibliography}{999}
\bibitem{Bertsch} G. F. Bertsch and S. Das Gupta, Phys. Rep. {\bf 160}, 4 (1988)
\bibitem{Bertsch2} G. F. Bertsch, H. Cruse and S. Das Gupta, Phys. Rev {\bf C 29}, 673 (1984)
\bibitem{Bauer} W. Bauer,G. F. Bertsch,S Das Gupta, Phys. Rev. Lett. {\bf 58}, 863 (1987)
\bibitem{Aichelin1} J. Aichelin and H. Stocker, Phys. Lett {\bf B 176}, 14 (1986)
\bibitem{Beauvais} G. E. Beauvais, D. H. Boal and J. C. K Wong, Phys Rev {\bf C 35}, 545 (1987)
\bibitem{Ayik} S. Ayik and C. Gregoire, Nucl. Phys. {\bf A 513}, 187 (1990)
\bibitem{Randrup} J. Randrup and B. Remaud, Nucl. Phys {\bf A 514}, 339 (1990)
\bibitem{Chomaz} Ph. Chomaz,G. F. Burgio and J. Randrup, Phys. Lett. {\bf B 254}, 340 (1991)
\bibitem{Rizzo} J. Rizzo,Ph. Chomaz and M. Colonna, Nucl. Phys. {\bf A 806}, 40 (2008)
\bibitem{Napolitani} P. Napolitani and M. Colonna, Phys. Lett {\bf B 726}, 382 (2013)
\bibitem{Cugnon} J. Cugnon,T. Mizutani and J. Vandermeulen, Nucl. Phys. {\bf A 352} (1981) 505
\bibitem{Lenk} R. J. Lenk and V. R. Pandharipande, Phys. Rev. {\bf C 39}, 2242 (1989).
\bibitem{Dasgupta} S. Das Gupta and A. Z. Mekjian, Phys. Rev {\bf C 57}, 1361 (1998)
\bibitem{Das} C. B. Das,S. Das Gupta,W. G. Lynch,A. Z. Mekjian and M. B. Tsang, Phys. Rep. {\bf 406}, 1 (2005)
\bibitem{Xu} H. S. Xu et al., Phys. Rev. Lett {\bf 85}, 716 (2000)
\bibitem{Frankland} J. D. Frankland et al., Nucl. Phys. {\bf A 649}, (2001) 940
\bibitem{Mallik11} S. Mallik, G. Chaudhuri and S. Das Gupta, arXiv:1503.04929, Submitted in Phys. Rev. {\bf C}.
\bibitem{Chaudhuri1} G. Chaudhuri,S. Das Gupta and M. Sutton, Phys. Rev. {\bf B 74}, 174106 (2006)
\bibitem{Chaudhuri2} G. Chaudhuri and S. Das Gupta, Phys. Rev. {\bf C 80}, 044609 (2009)
\bibitem{Bondorf} J. P. Bondorf, A. S. Botvina, A. S. Iljinv, I. N. Mishustin and K. Sneppen, Phys. Rep {\bf 257}, 133 (1995)
\bibitem{Botvina} A. Botvina, G. Chaudhuri, S. Das Gupta, I. Mishustin, Phys. Lett {\bf B 668} 414 (2008)
\bibitem{Souza} S. R. Souza et al., Nucl. Phys. {\bf A 571}, (1994) 159
\bibitem{Leray} S. Leray et al., Nucl. Phys. {\bf A 531}, (1991) 177
\bibitem{Aichelin} J. Aichelin and G. F. Bertsch, Phys. Rev {\bf C 31},1730 (1985)
\end{thebibliography}
\end{document}